\documentclass[  superscriptaddress, pra, twocolumn]{revtex4}

\usepackage{amsmath,amssymb,amsthm}
\usepackage{xspace}
\usepackage{graphicx} %for graphics
\usepackage{latexsym} %for \longrightarrow
\usepackage[mathscr]{eucal} %for \mathscr{}
\usepackage{bm}
\usepackage{dcolumn}
\usepackage{array}
\usepackage[usenames,dvipsnames]{color}

\newcommand{\floor}[1]{\lfloor #1 \rfloor}
\newcommand{\bra}[1]{\langle #1|}
\newcommand{\ket}[1]{|#1\rangle}

\begin{document}
\title{Delayed Choice Contextuality: A way to rule out Contextual Hidden Variables}

\author{Jayne \surname{Thompson}}
\affiliation{Centre for Quantum Technologies, National University of Singapore, 3 Science Drive 2, 117543 Singapore, Singapore}

\author{Mile \surname{Gu}}
\email{cqtmileg@nus.edu.sg}
\affiliation{Center for Quantum Information, Institute for Interdisciplinary Information Sciences, Tsinghua University, 100084 Beijing, China}
\affiliation{Centre for Quantum Technologies, National University of Singapore, 3 Science Drive 2, 117543 Singapore, Singapore}

\author{Pawe{\l} \surname{Kurzy\'nski}}
\affiliation{Centre for Quantum Technologies, National University of Singapore, 3 Science Drive 2, 117543 Singapore, Singapore}
\affiliation{Faculty of Physics, Adam Mickiewicz University, Umultowska 85, 61-614 Pozna\'{n}, Poland}

\author{Su-Yong \surname{Lee}}
\affiliation{Centre for Quantum Technologies, National University of Singapore, 3 Science Drive 2, 117543 Singapore, Singapore}

\author{Dagomir \surname{Kaszlikowski}}
\email{phykd@nus.edu.sg}
\affiliation{Centre for Quantum Technologies, National University of Singapore, 3 Science Drive 2, 117543 Singapore, Singapore}
\affiliation{Department of Physics, National University of Singapore, 2 Science Drive 3, 117542 Singapore, Singapore}

\pacs{03.67.-a, 03.65.Ta, 03.65.Ud}
\begin{abstract}
A PhD student is locked inside a box, imitating a quantum system by mimicking the measurement statistics of any viable observable nominated by external observers. Inside a second box lies a genuine quantum system. Either box can be used to pass a test for contextuality - and from the perspective of an external observer, be operationally indistinguishable. There is no way to discriminate between the two boxes based on the output statistics of any contextuality test. This poses a serious problem for contextuality tests to be used as viable tests for device independent quantumness, and severely limits realistic use of contextuality as an operational resource. Here we rectify this problem by building experimental techniques for distinguishing a contextual system that is genuinely quantum, and one that mimics it through clever use of hidden variables.
\end{abstract}

\maketitle
\section{Introduction}
Bell nonlocality is one of the most iconic and paradoxical aspects of quantum phenomenology. The realities of two space-like separated systems can be entangled, such that the choice of whether to measure $A_2$ or $A_3$ on one system can affect the measurement statistics of $A_1$ on the other \cite{Bell}. Contextuality generalized these ideas, noting that the essence of non-locality remains in single, localized quantum systems. The outcome of an observable $A_1$ can depend on whether it is comeasured with $A_2$ or $A_3$, even if neither observable $A_2$ nor $A_3$ disturbs the measurement statistics of $A_1$ \cite{Kochen}.

Unlike nonlocality, contextuality can be exhibited by classical as well as quantum systems \cite{Dakic,Harrigan,Kleinmann}. All contextuality experiments so far, have relied on collecting output statistics from a series of joint measurements $A_1 A_2$, $A_2 A_3$, etc. on some quantum state. While this approach has been immensely successful \cite{Zeilinger,Ahrens,Ambrosio,Zhang}, all outcomes in these tests can be replicated by classical systems. Suppose two black boxes are tested for contextuality. One contains a genuine quantum system that faithfully outputs the required quantum measurement statistics, the other contains a classical computer which merely simulates the mathematics of quantum theory. Current experimental tests of contextuality cannot distinguish which of the two is quantum. This stands in stark contrast to tests of nonlocality where local hidden variables can be excluded due to space like separation of Alice and Bob's choice of measurement settings and no-signalling. The inability to exclude hidden variables is a key limitation of current contextuality experiments, which in turn, hinders their use as a resource in device independent scenarios - such as generating certified random numbers  \cite{Kim}.

This article aims to mitigate this limitation. We introduce experimental methods to exclude contextual hidden variables through physical principles and apply them to refine experimental tests of contextuality in quantum systems. Our approach rests on the use of an external quantum mechanical degree of freedom. In synthesizing superpositions of different contextuality measurements conditioned on this degree of freedom, we can imprint the predictions of contextuality into correlations between the system of interest, and our quantum mechanical ancilla. The outcomes of contextuality experiments can thus be translated into a bipartite setting. In doing so, we can replace the standard ad hock assumption in contextuality tests - that any participating classical system did not reconfigure its answers based on the measurement settings  - with a much more physically motivated assumption of no-signaling. This puts tests of contextuality on a similar footing to nonlocality tests, and ultimately reduces the current gap between them.

The article is organized as follows. Section \ref{sec:contextualitytests} will introduce the formal framework of contextuality. Section \ref{sec:results} will outline the specifics of our proposed experiment, together with corresponding proofs of how it can distinguish contextual hidden variables from genuine quantum contextuality.  Section \ref{sec:blackbox} describes these ideas within the general Black Box framework. Section  \ref{sec:discusion} then concludes the paper with discussions.

\section{Background and Definitions}\label{sec:contextualitytests}
Suppose that a system has $n$-observables $\mathcal{M}_A = \{A_0, \dots, A_{n-1}\}$, including an observable $A_i$ which is compatible with several distinct proper subsets of $\mathcal{M}_A$. Each compatible subset constitutes a measurement context for $A_i$.  The non-contextual hypothesis is: the outcome of $A_i$ is independent of which subset you choose to measure $A_i$ alongside. Quantum systems with three or more Hilbert space dimensions violate his hypothesis - they are innately contextual \cite{Kochen}.

Tests are formulated in terms of non-contextual inequalities; mathematical inequalities that constrain all non-contextual systems. An archetypical example is the $n$-cycle inequality:
\begin{equation}\label{eq:ncycles}
\sum_{i=0}^{n-2} \langle A_i A_{i+1}\rangle  + (-1)^{n-1} \langle A_{n-1} A_{0}\rangle \ge 2-n,
\end{equation}
where observables $A_0, \dots A_{n-1}$  have dichotomic $\pm 1$- outcomes, all addition is done modulo $n$, and any consecutive pair $A_i$ and $A_{i+1}$ can be measured simultaneously \cite{Braunstein,Araujo,Klyachko}. This inequality is derived under the no-disturbance assumption --  the marginal distributions $p(a_j |A_j)$ is assumed to be independent of the measurement context, i.e. $p(a_i|A_i) = \sum_{a_{i+1}}p(a_i a_{i+1}|A_i A_{i+1}) = \sum_{a_{i-1}}p(a_i a_{i-1}|A_i A_{i-1})$. Provided this assumption is upheld, any non-contextual hidden variable will satisfy \eqref{eq:ncycles}. Meanwhile quantum systems can violate \eqref{eq:ncycles} up to the Tsirelson bound  ${\mathcal T}_n = -n\cos{\pi/n}$ if $n$ is even, or ${\mathcal T}_n = (n - 3n\cos{\pi/n})/(1 + \cos{\pi/n})$ if $n$ is odd \cite{Araujo,Tsirelson}.

Contextual hidden variables can also violate this inequality. Consider a hidden variable that flips an unbiased coin to generate $a_i = -1$ or $ +1$ at random, then computes $a_{i+1} = -a_i + 2 a_{n-1} (n \,\, {\rm mod} \,\,2)\delta_{i, n-1}$. This hidden variable will violate the $n$-cycle inequality up to the arithmetic bound of $-n$. Thus contextuality is not unique to quantum systems. It is also exhibited by classical processes which take the full complement of observables being simultaneously measured $\{A_i, A_j, \dots\}$ and uses this input to generate a context-dependent-outcome $a_i$ for each observable $A_i$, i.e. a contextual hidden variable.

This observation limits the capacity of standard contextuality tests to guarantee the non-classicality of an untrusted physical system. An unknown system may violate a noncontextual inequality simply by hosting a computer that executes a suitable contextual hidden variable model. Indeed many such automata have already been proposed \cite{Kleinmann,Dakic,Harrigan}.

\textbf{Framework}. This motivates us to consider the following scenario. Alice wishes to test if a quantum system $\rho$ inside a black box is contextual, while excluding contextual hidden variables.  This may be motivated by a desire to generate quantum random numbers through a certifiable quantum source of contextuality. Her adversary wishes to cheat her by replacing this black-box with an imitation that mimics the measurement results through contextual hidden variables \cite{Kleinmann}. Alice is tasked with distinguishing the genuinely quantum system from the imitation and simultaneously checking for contextuality - i.e. the task of discriminating between these two boxes is equivalent to excluding a contextual hidden variable explanations for her experimental data.

Alice has an experimental apparatus, which can be pre-configured to measure any pair of observables $A_{i}$ and $A_{i+1}$. She also has a source of black boxes; supplied by an untrusted third party with the capability to generate classical or quantum boxes. In each run Alice uses the source to generate a  black box, and then inserts it into the experimental apparatus, without knowing whether it contains a quantum or classical system. The apparatus then returns two measurement outcomes $q_0$ and $q_1$, respectively. At the end of the run Alice uses the outcomes as data points for $a_i$ and $a_{i+1}$ in her experiment. We note two properties that are required for the experimental test to conclude a system is contextual:
\begin{itemize}
\item[(i)]{If an observable $A_i$ is measured twice during a single run, then both answers should be the same. \label{item1}}
\item[(ii)]{The data will violate the $n$-cycle inequality in Eqn.\eqref{eq:ncycles}.~\label{item2}}
\end{itemize}

Alice will thus screen her data for compliance with both (i) and (ii).  Criterion (i) is often applied to strings of measurements $A_i, A_{i+1}, A_i$. Checking the outcomes of $A_i$ (and of $A_{i+1}$) are consistent in any string of measurements $A_i A_{i+1} A_i\dots$, is an operational way of verifying the no-disturbance assumption. This check is often advocated and used in experimental implementations, where the emphasis is on checking the quantum measurements commute.

In the following sections, we develop experimental methods for excluding contextual hidden variables and use them to refine existing experimental tests of contextuality. Our new refined tests can exclude contextual hidden variables, due to inability to satisfy the aforementioned criteria (i) and (ii).

\section{The Protocol}\label{sec:results}

To implement the aforementioned standard contextuality tests, Alice needs to choose measurement settings at random. This involves the use of a reliable source of randomness. One method, for example, is via the use of an ancillary qubit $\mathcal{B}$ initiated in state $\ket{+} = \frac{1}{\sqrt{2}}\ket{0} + \frac{1}{\sqrt{2}}\ket{1}$. By measuring this qubit in the standard Pauli $Z$ basis (i.e., the $\ket{0}, \ket{1}$ basis), Alice can obtain a random bit $b = \{0,1\}$ that determines which measurement she will perform. For example, Alice may use the random bit to decide which of two sets of mutually commuting observables  $S = \{A_{S1},A_{S2},\dots\}$, or $S' = \{A_{S'1}, A_{S'2},\dots \}$, where $ S,S' \subset \mathcal{M}_A$ to measure. That is she configures her apparatus to co-measure $A_{S1}, A_{S2},\dots$ if the random bit if $b = 0$,  and  $A_{S'1}, A_{S'2},\dots $ if $b = 1$.

\begin{figure}[Htb]
\centering
\includegraphics[width=0.5\textwidth]{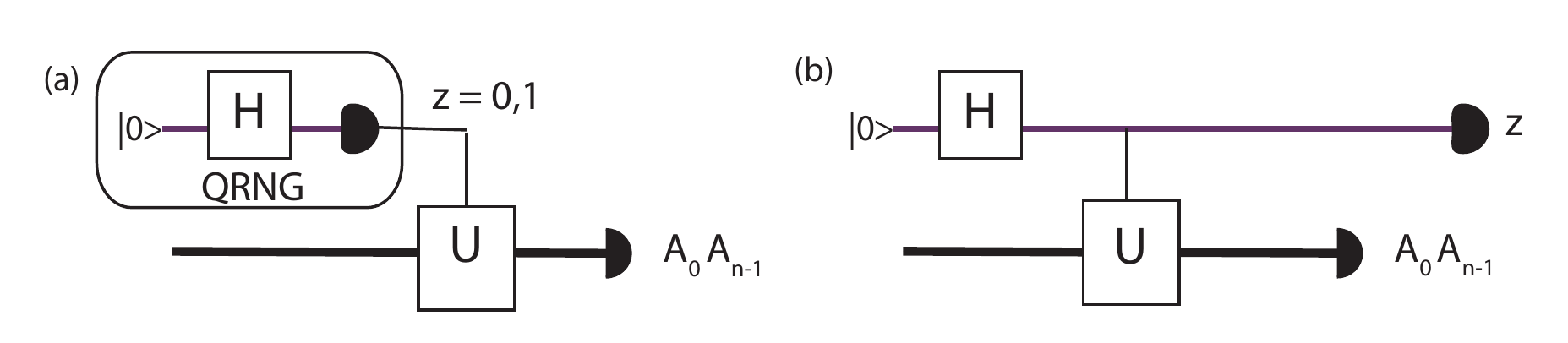}
\caption{Figure (a) depicts a classical control scheme used in standard contextuality tests, where the $Z$ basis measurement outcomes of an ancillary qubit $\rho_c = \ket{+}\bra{+}$, are used as a quantum random number generator (QRNG). If the QRNG outputs 0 then the observables $S =\{A_0, A_{n-1}\}$ are measured, else when the QRNG outputs 1 the measurement settings are flipped to $S' =\{A_j,A_{j+1}\}$. In the revamped protocol (b) a control-unitary is used to reverse the order of measurement, so the settings can be flipped retroactively - that is we measure the system we are testing for contextuality, afterwards we fix the settings by measuring the ancilla in the $Z$ basis. This figure highlights  the tight knit relationship with quantum delayed choice protocols \cite{Terno}.  \label{fig:schematic}}
\end{figure}

Our proposal is based on reversing this order of events. Instead of pre-selecting the measurement setting, Alice will delay this choice until after the system of interest $\mathcal{A}$, has been measured. Formally,  suppose there exists a unitary $U$, that transforms each element of $S$ to an element of $S'$, i.e., $S' = \{U^{\dagger} A_{S1} U, U^{\dagger} A_{S2} U, \dots\}$. For $n$-cycle tests based upon the standard observables ${\cal M}_A$ as in \cite{Araujo,Braunstein,Grudka}, the compatible subsets $S\subseteq \{A_j, A_{j+1}\}$ and $S' \subseteq\{A_i, A_{i+1}\}$ where $i,j \in \{0,\dots, n-1\}$, satisfy this assumption -- provided we ensure $S$ and $S'$ have the same number of elements.

Consider the following general strategy. Let $\rho$ represent the state of the system of interest, and $\rho_c$ the state of the ancilla qubit. Instead of measuring the system directly, Alice first applies the aforementioned $U$ controlled on the $Z$ basis of the ancilla qubit, i.e., Alice applies $C_U: \ket{b}\ket{\phi} \rightarrow U^b \ket{b}\ket{\phi}$. Following this process, Alice makes the standard measurements associated with each of the observables in $S$ to obtain outcomes $q_0,q_1,\dots$. Once these measurement outcomes are obtained, Alice generates the random number $b \in \{0,1\}$ by a $Z$ measurement of the ancilla (See Fig. \ref{fig:schematic} for details). When $b= 0$, Alice treats $q_0,q_1,\dots$ as the outputs of measuring the elements of $S$ and use them to collect statistics on $\langle A_{S1} A_{S2}\rangle$. Otherwise the outputs are treated as if the elements of $S'$ were measured; and used to  evaluate $\langle A_{S'1}A_{S'2}\rangle$ for elements of $S'$.

Thus, Alice now decides what questions to ask a possibly contextual system of interest \emph{after} she receives the answers to the said questions. The main idea is that the black box's incumbent PhD student must now generate outcomes, without knowing exactly what observables are being measured. Such a PhD student can no longer execute a contextual hidden variable strategy, which uses the full complement of observables being measured to compute outcomes. In what follows, we demonstrate appropriate choice of $S$ and $S'$ allows Alice to exclude classical contextual hidden variables.  The exact specifics will vary depending on whether the number of observables, $n$, is even or odd.

\label{sec:evenncycles}
\textbf{When $n$ is even}, we execute the protocol above with $S = \{A_0, A_{n-1}\}$  and $S' = \{A_j, A_{j+1}\}$ for some $j \in\{0,\dots,n-2\}$. In instances where the ancilla measurement outcome $b$, is $0$, we have effectively measured $A_0$ and $A_{n-1}$ and thus can collect statistics about $\langle A_0 A_{n-1}\rangle$. In instances where $b = 1$, similar reasoning allows us to collect information on $\langle A_{j} A_{j+1}\rangle$. For example, repetition of the protocol when $j$ is set to $0$ allows estimation of $\langle A_0 A_{n-1}\rangle$ and $\langle A_0 A_1\rangle$. Reiterating this procedure for $j=1,\dots,n-2$ each time allows evaluation of each of the correlation terms in Eq. \eqref{eq:ncycles}.

Consider a contextual classical box that attempts to mimic these statistics. Without loss of generality, we can model such a box by assuming it contains an internal memory - a contextual hidden variable $\lambda$. The box is assumed to have complete knowledge of Alice's experimental setup (i.e, what $S$ and $S' =\{A_j A_{j+1}\}$ are), but no knowledge about the outcome of Alice's eventual measurement of the ancilla (we will see in Section \ref{sec:blackbox} that the latter assumption can be verified experimentally).

In each run Alice makes a measurement for each observable in $S=\{A_0,A_{n-1}\}$. To successfully fool Alice, the box must replicate the  quantum output statistics for $q_0$ and $q_1$. In the following we show that this is impossible; a classical box with contextual hidden variables cannot consistently satisfy conditions (i) and (ii) from Sec. \ref{sec:contextualitytests}, making it impossible for such a system to mimic genuine quantum contextuality without detection.

\textit{Proof of Result.} In order to violate the $n$-cycle inequality \eqref{eq:ncycles}, it is necessary that
\begin{equation}\label{eq:inequality}
C_j =\langle A_j A_{j+1} \rangle - \langle A_{n-1} A_0\rangle  < 0.
\end{equation}
This holds independently for each  $j \in \{0,\dots, n-2\}$. To see this write  \eqref{eq:ncycles} as $ C_j + D_j \ge -n+2$, where $ D_j = \sum_{i=0}^{n-2} \langle A_i A_{i+1}\rangle - \langle A_j A_{j+1}\rangle $. Because $D_j$ contain $n-2$ correlation terms which satisfy $-1\le \langle A_i A_{i+1}\rangle \le1$, we are guaranteed that $ D_j \ge 2-n  $. Hence in order to violate  \eqref{eq:ncycles}  we need $C_j<0$.

We take the outcomes $q_0, q_1$, collected from runs where $S=\{A_0, A_{n-1}\}$ and $S' = \{A_j A_{j+1}\}$, in conjunction with the output of the ancilla measurement, $b$, and evaluate
\begin{eqnarray}\label{eq:correlations}
&& C_j = p(q_0 = q_1|b =1) - p(q_0 \neq q_1|b =1)  \notag \\
&&  -  p(q_0 = q_1|b =0) + p(q_0 \neq q_1|b=0).
\end{eqnarray}
We observe that if we define probability distributions $P(q_0 q_1) = p(q_0 q_1| b = 1)$ and $Q(q_0 q_1) = p(q_0 q_1| b = 0)$, then
\begin{eqnarray}\label{eq:variationaldistance}
|C_j| \le \sum_{q_0, q_1}| P(q_0 q_1) - Q(q_0 q_1)|= D(P,Q)
\end{eqnarray}
where $D(P,Q)$ is the variational distance between $P(q_0q_1)$ and $Q(q_0q_1)$.

Let us assume a contextual hidden variable $\lambda$ supplied the outcomes $q_0$ and $q_1$. This assignment must be made without knowing $b$, the future outcome of measuring the ancilla qubit in the $Z$ basis. We note that $b$ equiprobable to be 0 or 1, since the ancilla is prepared in a $\ket{+}$ state and the control unitary does not effect the $Z$ statistics of the ancilla (i.e. the control unitary commutes with a $Z$ basis measurement of the ancilla, since with respect to the $Z$ basis of the ancilla the control unitary is $I \oplus U$, where $\oplus$ is the direct sum and $I$ is the identity.). With no prior knowledge of $b$ available when $\lambda$ is assigned $p(\lambda | b) = p(\lambda)$, and
\begin{eqnarray}\label{eq:probdist}
&&p(q_0 q_1 z_c ) = \int p(q_0 q_1| \lambda)p(\lambda|b) p(b) d\lambda  \\ &&= p(b) \int p(q_0 q_1| \lambda)p(\lambda) d\lambda = p(b) \times p(q_0 q_1)\notag
\end{eqnarray}
hence the variational distance $D(P,Q)$ in Eq. \eqref{eq:variationaldistance} is precisely 0. A contextual hidden variable can not violate Inequality \eqref{eq:inequality} and therefore will not violate the $n$-cycle inequality \eqref{eq:ncycles}.

Of course if the experiment is not repeated enough times, then a classical system could appear to satisfy \eqref{eq:inequality} due to random statistical fluctuations. In practice this problem effects all contextuality tests, and has a standard resolution. Quantum systems actually violate the $n$-cycle inequality up to the Tsirleson \cite{Tsirelson} bound $\mathcal{T}_n$ in Sec. \ref{sec:contextualitytests} -- so the quantum statistics ideally satisfy a stricter inequality $C_j \le {\mathcal T}_n +(n-2)$. Alice will pick an $\epsilon$ satisfying $0 < \epsilon < |{\mathcal T}_n +(n-2)|$, and check if the experimental outcomes satisfy:
\begin{equation}\label{eq:epsilonineq}
\langle A_j A_{j+1} \rangle - \langle A_{n-1} A_0\rangle  < -\epsilon.
\end{equation}
Due to the central limit theorem, the probability that statistics generated by a classical hidden variable, will satisfy \eqref{eq:epsilonineq}, dies off exponentially with the number of runs ${\cal N}$.

\begin{figure}[Htb]
\centering
\includegraphics[width=0.3\textwidth]{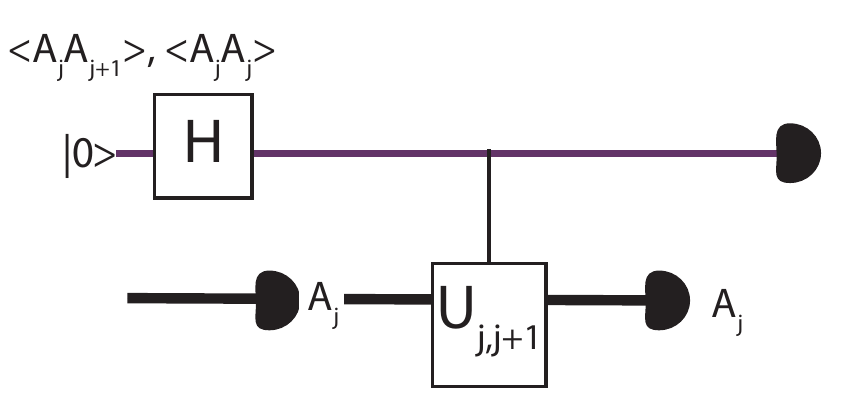}
\caption{In the odd $n$ cycle case we use an alternate protocol, where each run allows us to measure $\langle A_j A_{j+1}\rangle$ for some $j\in\{0,\dots, n-1\}$, or $\langle A_j A_j\rangle$. A non-destructive measurement of the observable $A_j$ on the register is followed by a control unitary $C_{U_{j,j+1}}$ where $U_{j,j+1}^{\dagger} A_j U_{j,j+1} = A_{j+1}$. Finally we repeat the measurement of $A_j$, and measure the ancilla in the $Z$ basis generating an outcome $b\in\{0,1\}$. Supposing we start with $j=0$ then post selecting on runs where the ancilla outcome $b= 1$, we can evaluate $\langle A_0 A_1\rangle$ meanwhile when $b= 0$ the data yields $\langle A_0 A_0\rangle$. \label{fig3}}
\end{figure}

\textbf{When $n$ is odd}, The protocol is summarized by Fig. \ref{fig3}. Alice intends to evaluate the correlation term $\langle A_j A_{j+1}\rangle$ for some $j \in \{0,\dots, n-1\}$. The main difference here is that Alice first implements a non-destructive measurement of observable $A_j$ on system $\mathcal{A}$, with outcome $q_0$. Once done, she then implements the procedure in Sec. \ref{sec:results}, with $S=\{A_j\}$ and $S' = \{A_{j+1}\}$.
That is she introduces an ancillary qubit $\rho_c = \ket{+}\bra{+}$, and applies the unitary $U$ satisfying $U^{\dagger}SU = S'$ to $\mathcal{A}$, controlled on the $Z$ basis of $\rho_{c}$.

To determine what measurements she has effectively made, Alice completes the protocol by measuring the ancilla in the $Z$ basis, with outcome $b$. On runs where $b=0$, Alice attributes the outcomes $q_0,q_1$ to two consecutive measurements of $A_j$. Otherwise, Alice treats $q_0$ as an outcome of measuring $A_j$ and $q_1$ as an outcome of measuring $A_{j+1}$. Thus, depending on the outcome $b$, Alice can collect statistics about either $\langle A_j A_j\rangle$ or $\langle A_j A_{j+1}\rangle$.

Analogously to the even $n$-cycle case, a classical box with contextual hidden variables must replicate the output statistics of $q_0$ and $q_1$, for both possible measurement outcomes of the ancilla. This is not possible.

\textit{Proof of Result.} In order to satisfy the criteria (i) and (ii) from Sec. \ref{sec:contextualitytests}, it is necessary that
\begin{equation}\label{eq:cstar}
C_j^* =  \langle A_j A_{j+1} \rangle - \langle A_j A_{j}\rangle <0.
\end{equation}
 This inequality must be satisfied independently for each $j \in\{0,\dots, n-1\}$. This follows from observing two points. Firstly Criterion (i) implies $\langle A_j A_{j}\rangle = 1$. Secondly to violate the $n$-cycle inequality (Criterion (ii)), we require $\langle A_j A_{j-1}\rangle < 1$  for all $j \in \{0,\dots, n-1\}$. We can establish this using proof by contradiction: assume  $\langle A_j A_{j-1}\rangle = 1$ for some $j \in \{0,\dots n-1\}$ and we observe violation of Inequality  \eqref{eq:ncycles}. This would imply $\sum_{i=0}^{n-1}\langle A_i A_{i+1} \rangle - \langle A_j A_{j+1}\rangle < 1-n$. But this sum only contains $n-1$ terms each lower bounded by -1. This is a contradiction.

 By mirroring the analysis in the even case, specifically Eq. \eqref{eq:correlations}-\eqref{eq:probdist}, we observe
\begin{eqnarray}
&& C_j^* = p(q_0 = q_1|z =1) - p(q_0 \neq q_1|z =1)  \notag \\
&&  -  p(q_0 = q_1|z =0) + p(q_0 \neq q_1|z =0).
\end{eqnarray}
Again $|C_j^*| < D(Q,P)$, where $P(q_0 q_1)$ and $Q(q_0 q_1)$ are defined analogously to the even $n$ case. If $q_0$ and $q_1$ are generated by a contextual hidden variable $\lambda$, then using the same
argument as in Eq. \eqref{eq:probdist} we conclude $D(Q,P)$ is exactly zero provided no prior information about $b$ is available to influence the choice of $\lambda$. Hence the contextual box will not satisfy
\eqref{eq:cstar}, implying that either condition (i) or (ii) from Sec. \ref{sec:contextualitytests},  will be
contradicted.

\section{Experimental Technicalities}
As with all contextuality experiments, the physical implementation of this protocol must satisfy an important caveat.

During this implementation Alice will need to collect data for all $n$ correlation terms $\langle A_0 A_1\rangle, \,\langle A_1 A_2\rangle, \dots, \langle A_0 A_{n-1}\rangle$. For each correlation term Alice must set up her experiment in Figs. \ref{fig:schematic}-\ref{fig3} using a different $C_U$ gate. If an observable $A_j$ is measured in two different setups -- i.e. the setups for $\langle A_j A_{j-1}\rangle $ and $\langle A_j A_{j-1}\rangle$ -- then the actual measurement $A_j$ must be physically identical in both cases.  Otherwise any observed dependence of the outcomes $a_j$ on whether we measure $A_j A_{j-1}$ or $A_j A_{j+1}$ (i.e. any contextuality) could be due to imperfect reconstruction of $A_j$ when changing the settings.

\begin{figure}[Htb!]
\centering
\includegraphics[width=0.5\textwidth]{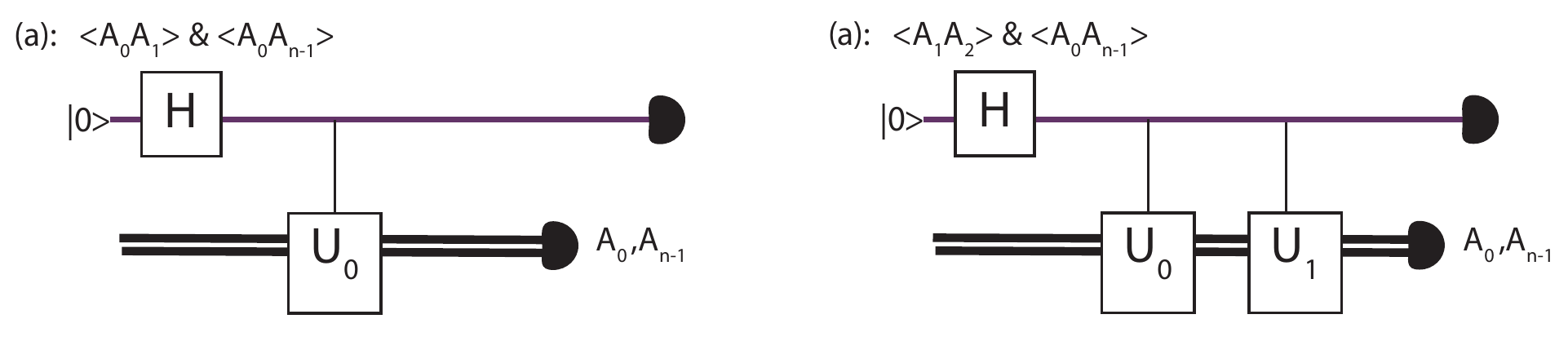}
\caption{The refined protocol applied to even $n$ cycles. Figure (a) demonstrates the example $j= 0$, which allows us to measure the correlation terms $\langle A_0 A_{n-1}\rangle $ and $\langle A_0 A_1\rangle $. Unitary $U_0$ is applied to the register, controlled on the state of the ancilla $\rho_c = \ket{+}\bra{+}$. Note $U_0$ is chosen so that it commutes with measurements of $A_0$, i.e. $U_0^{\dagger} A_0 U_0 = A_0$, and satisfies $U_0^{\dagger} A_{n-1}U_0= A_1$. Afterwards we measure $A_0$ and $A_{n-1}$ on the register, followed by a $Z$ basis measurement on the ancilla. In (b) we need to extend this protocol to get the next correlation term $\langle A_1 A_2\rangle$. However we must ensure that in both (a) and (b), the measurements associated with $A_0, A_{n-1}$ and $A_1$ are physically identical. To do this we follow the techniques in Ref.  \cite{Zeilinger}. We introduce a new control unitary $C_{U_1}$ such that $U_1^{\dagger} A_0 U_1 = A_2$ while $U_{1}^{\dagger} A_{1} U_{1} = A_1$, so that measurements of observable $A_1$  commute with the action of $C_{U_1}$. Note that measurements of $A_0,A_{n-1}$  will also commute with the $C_{U_1}$ gate (i.e. to measure $A_0 A_{n-1}$ we postselect on runs where the effect of the $C_{U_1}$ gate is the same as an identity transformation). In principle, because the only difference between (a) and (b) is a $C_{U_1}$ gate, and this gate does not effect the measurement outcome of $A_1$, or $A_0,A_{n-1}$, these observables should be the same in both setups. Furthermore the action of the two control unitaries is equivalent to a single control gate $C_{U_1 U_0}$ where $U_1^{\dagger} U_0^{\dagger} A_{n-1} U_0 U_1 = A_1$ while $U_1^{\dagger} U_0^{\dagger} A_0 U_0 U_1 = A_2$.  Iteratively repeating this method along with the techniques in Ref.  \cite{Zeilinger} allows us to acquire the remaining correlation terms $\langle A_2 A_3\rangle, \dots$ in the $n$-cycle inequality. \label{fig2}}
\end{figure}

In practice this problem is solved in standard contextuality tests by Ref.  \cite{Zeilinger} and others. We simply refine the contextuality test in Ref.  \cite{Zeilinger} to ensure we have also fulfilled this basic tenant. The even case is illustrated in Fig. \ref{fig2}, the odd case will follow the same principles and therefore has not been included here.

\section{The Black-Box Framework}\label{sec:blackbox}
So far we have made use of quantum mechanical terminology. This is in line with the paper's focus on describing concrete experiments that can rule out contextual hidden variables. Of course, contextuality is often discussed in a completely black-box setting - where the only properties of a system are its output statistics given input questions; and no assumptions are made that the system is quantum mechanical. Our proposal generalized naturally into this picture.

Because the security of this protocol relies on independence of the control from the system being tested for contextuality; we need to introduce two parties Alice and Bob, who share a pair of black boxes. Indeed, since the quantum protocol required Alice and Bob to share $n$-different entangled resource states -- one for each of Alice's settings choices $j=0,\dots, n-1$, in the $n$-cycle test --  the black box analogue actually requires  $n$ pairs of boxes, labeled $j=0,\dots, n-1$. We want recast the measurements in Figs. \ref{fig:schematic}-\ref{fig2} as a series of questions Alice and Bob ask their respective boxes.

In each round of the $n$-cycle test Alice (and Bob) now select a pair of boxes labeled `$j$'. Alice asks her box two sequential dichotomic questions $Q_0$ and $Q_1$ and gets answers $q_0, q_1 = \pm 1$. Afterwards Bob asks his box a single question $Z$ which generates outcome $b =0$ or $1$. The outcomes $q_0,q_1,b$ can be treated as the outcome from Sec. \ref{sec:results} for the aforementioned values of $j$ and $n$.

\textbf{Verifying non-determinism of measurement selection.} Recall that our proofs above relied on the crucial assumption that the box does not have access to the value of $b$ prior to outputting $q_0$ and $q_1$. This relies of the assertion that there is no hidden variable that determines the value of $b$ prior to its measurement. The black-box framework gives us a natural way to explicitly verify this assumption.

On some runs, after selecting their boxes Alice and Bob may randomly choose to perform a Bell test between their box pairs. During this Bell test Alice chooses to ask her box one of two possible questions $\bar{A}_1$ or $\bar{A}_2$; while Bob chooses to asks his box either question $\bar{B}_1$
or $\bar{B}_2$. $\bar{B}_1$ is selected to be the same question, $Z$, Bob would have asked in delayed choice contextuality. The resulting statistics are used to test violation of the CHSH inequality \cite{Clauser}:
\begin{equation}\label{eq:chsh}
\langle \bar{B}_1 \bar{A}_1 \rangle +\langle \bar{B}_2 \bar{A}_2\rangle +\langle \bar{B}_2 \bar{A}_1\rangle -\langle \bar{B}_1 \bar{A}_2\rangle \le 2.
\end{equation}
The violation of this inequality certifies there is no realistic description for outcomes of question $\bar{B}_1$. This demonstrates that there is no hidden variable that predicts Alice's decision of whether to ask questions in $S$ or $S'$. Thus any capacity for the Box to gain this knowledge will violate causality.

In the specific case of quantum systems, the Bell test is performed on the final state of the circuits in Figs. \ref{fig2}-\ref{fig3}. The control qubit is assumed to be inside Bob's box $j$; and the register inside Alice's box `$j$'. $\bar{B}_1$ represents a $Z$-measurement, in line with our constraint that it asks the same question as our refined contextuality test. Provided the register was in some pure state $\ket{\chi}$ directly before the control unitary gate $C_U$, the resulting bipartite system is able to violate the CHSH inequality for any $n$ for some choices of $\bar{B}_2$, $\bar{A}_1$ and $\bar{A}_2$.

Specifically let $\bar{B}_2$ be a $X$ measurement; such that Bob measures the ancilla in either the $X$ or $Z$ basis. Meanwhile define $\ket{\bar{0}} = \ket{\chi}$, $\ket{\bar{1}} = \frac{U\ket{\chi} - \sqrt{a} \ket{\chi}}{ \sqrt{1-a}}$ where $\sqrt{a} =\bra{\chi}U\ket{\chi}$, as an effective qubit on Alice's system of interest; with associated Pauli operators $\Sigma_x = \ket{\bar{0}}\bra{\bar{1}} +\ket{\bar{1}}\bra{\bar{0}}$ and $\Sigma_z = \ket{\bar{0}}\bra{\bar{0}} - \ket{\bar{1}}\bra{\bar{1}}$; and let $\bar{A}_1 = \sin{\theta} \,\Sigma_z + \cos{\theta} \, \Sigma_x$ and $\bar{A}_2 = \sin{\phi}\, \Sigma_x + \cos{\phi}\,\Sigma_z$ represent the two measurement settings for Alice. Application of these measurement settings allows violation of the CHSH inequality up to $1+\sqrt{a} +\sqrt{2-a}-a > 2$, for all $a <1$   . For more details see the appendices.

\section{Discussion}\label{sec:discusion}

Here we have developed experimental techniques for excluding contextual hidden variables and applied them to refine contextuality experiments. By controlling the measurement settings in the contextuality test on an external ancillary quantum mechanical degree of freedom, we present a method for effectively concealing which observable is being measured in the contextuality test. This will effectively thwart a contextual hidden variable that needs to use prior knowledge of which observables are being measured in order to preassign a contextual outcome.

One of the major stumbling blocks in the use of contextuality as a quantum resource in line with non-locality has been its capacity to be simulated by classical information processing. How can we use contextuality as a resource if we can replace the same system with a student in a box? In developing an experimental method to certify a quantum state is contextual, and simultaneously exclude contextual hidden variables, our protocol helps address this problem. This innovation may thus lead to potential applications for contextual quantum systems - the natural candidate being, for example, the certified generation of random numbers \cite{Renner,Pironio,Grudka,Kim}

It would be interesting to see if some of the other problems surrounding contextuality tests could be addressed using similar ideas. For instance the no-disturbance assumption -- which is the direct equivalent of no-signalling in Bell tests -- is a hot button topic in contextuality tests. A key issue is that hidden variable models which do not satisfy the no-disturbance condition  (i.e. the marginal distribution they predict $p(a_i |A_i) = \sum_{a_{j}} p(a_i a_{j} |A_i A_j)$ depends on the measurement context $A_j$) can violate a noncontextual inequality. This opens an interesting possibility: could  a refined protocols, which effectively hides the measurement context `-' in any joint measurement $A_i $`-' , help address this problem? Furthermore the topic of contextuality as a resource is controversial, particularly since according to state independent contextuality protocols even the completely mixed state is contextual. By imprinting the predictions of contextuality into correlations between the test system and an external ancilla, we give a new interpretation for contextual statistics, which also links back to the quantum circuit formalism.  This leaves a host of interesting problems which could be addressed in future work.

Another point of interest, is the close relationship between our protocol and quantum delayed choice, a class of quantum protocols which have been experimentally demonstrated  in the context of optical interferometry~\cite{Terno,Celeri,Ionicioiu,Mann,Obrien,Tang,Ma,Coles}. The measurement statistics of a photon exiting an interferometer depend on a setting in the interferometer which can be either open and closed. By controlling whether the interferometer is open or closed on an ancillary quantum degree of freedom, we can effectively toggle the interferometer setting between open and closed after the photon has been measured  \cite{Terno}. This innovation, provides a remarkable way to rule out a hidden variable description for the  photon's measurement statistics. While the principles involved are similar to Wheeler's delayed choice \cite{Wheeler,Wheeler2}, quantum delayed choice is known to be more versatile and have applications to ruling out hidden variables in more general settings such as Bell tests \cite{Celeri}. Quantum delayed choice also furnishes a technique for measuring complementary phenomena using a single configuration of the measurement apparatus \cite{Terno}. Our work highlights the natural confluence of these ideas with tests of quantum contextuality.

\section{Acknowledgements}
J.T. would like to acknowledge input and discussions with Marek Wajs, Kavan Modi, Daniel Terno and Kihwan Kim.  This project/publication was made possible through the support of the John Templeton Foundation (grant number 54914) and the Foundational Questions Institute. The opinions expressed in this publication are those of the author(s) and do not necessarily reflect the views of the John Templeton Foundation.  The project was also supported by the National Research Foundation and the Ministry of Education in Singapore through the Tier 3  MOE2012-T3-1-009 Grant ``Random numbers from quantum processes" as well as the National Basic Research Program of China Grant 2011CBA00300, 2011CBA00302, the National Natural Science Foundation of China Grants 11450110058, 61033001, and 61361136003.

\section{Appendix A: Auxiliary Bell Test}
Here we elaborate on the Bell test in more detail. We illustrate the idealised scenario where the control qubit is in a state $\rho_c = \ket{+}\bra{+}$ directly before the control unitary gate, while the register is in a pure state $\rho =\ket{\chi}\bra{\chi}$. In standard $n$-cycle experiments the system being tested for contextuality is pure which should substantiate this assumption, see for instance \cite{Klyachko,Zeilinger,Braunstein}. However any additional noise in the system being tested for contextuality should simply decrease the predicted Bell violation. Given these assumptions the state of the circuit after the control unitary $C_U$ is:

\begin{eqnarray} \label{eq:state}
\ket{cir}&=& \frac{1}{\sqrt{2}}\ket{0}\ket{\chi} +\frac{1}{\sqrt{2}} \ket{1} U \ket{\chi}.
\end{eqnarray}

 We use the two subspaces $\ket{\bar{0}} = \ket{\chi}$, $\ket{\bar{1}} = \frac{U\ket{\chi} - \sqrt{a} \ket{\chi}}{ \sqrt{1-a}}$ where $\sqrt{a} =\bra{\chi}U\ket{\chi}$, to define an effective qubit on Alice's system of interest.  We then equip Alice with two observables $\bar{A}_1 = \sin{\theta} \,\Sigma_z + \cos{\theta} \, \Sigma_x$ and $\bar{A}_2 = \sin{\phi}\, \Sigma_x + \cos{\phi}\,\Sigma_z$ defined interms of  Pauli operators for her effective qubit  $\Sigma_x = \ket{\bar{0}}\bra{\bar{1}} +\ket{\bar{1}}\bra{\bar{0}}$ and $\Sigma_z = \ket{\bar{0}}\bra{\bar{0}} - \ket{\bar{1}}\bra{\bar{1}}$. And re-expresses the circuit's final state as
\begin{equation}\label{eq:circstate}
 \ket{\overline{circ}} = \frac{1}{\sqrt{2}}\ket{0}\ket{\bar{0}} +\sqrt{\frac{a}{2}} \ket{1}  \ket{\bar{0}} +\sqrt{\frac{1-a}{2}}\ket{1} \ket{\bar{1}}.
 \end{equation}
 Let's assume the control qubit goes to Bob and the register to Alice; and that Bob uses with the standard Pauli observables $\bar{B}_2 = X$ and $\bar{B}_1 = Z$ on his qubit.
 Now Alice and Bob can test a Bell inequality:
\begin{equation}
\langle \bar{B}_1 \bar{A}_1 \rangle +\langle \bar{B}_2 \bar{A}_2\rangle +\langle \bar{B}_2 \bar{A}_1\rangle -\langle \bar{B}_1 \bar{A}_2\rangle \le 2.
\end{equation}
In terms of the condensed notation $CHSH = \bar{B}_1 \bar{A}_1 + \bar{B}_2 \bar{A}_2 + \bar{B}_2 \bar{A}_1 - \bar{B}_1 \bar{A}_2$, the expectation value on state \eqref{eq:circstate} is:
\begin{eqnarray}\label{eq:chshexpectation}
 \langle CHSH \rangle &&= \left(1 +\sqrt{a}- a\right)\sin{\theta} +\sqrt{1-a} \left(1-\sqrt{a}\right) \cos{\theta} - \notag\\ &&\left(1-\sqrt{a}-a\right) \cos{\phi} +\sqrt{1-a}\left(1+\sqrt{a}\right)\sin{\phi}
\end{eqnarray}
 To find the best possible Bell violation $\langle CHSH\rangle_{max}$, we optimize \eqref{eq:chshexpectation} as a function of $\theta$, $\phi$. We know this quantity is greater than (or equal to) the value of \eqref{eq:chshexpectation} when $\theta = \pi/2$, i.e.:
\begin{eqnarray}
\langle CHSH \rangle_{max} &&\ge \left(1+\sqrt{a}-a\right) - \left(1-\sqrt{a} -a\right) \cos{\phi} \notag\\ &&+ \sqrt{1-a} \left(1+\sqrt{a}\right)\sin{\phi}
\end{eqnarray}
Now we can use $\sin{\left(\phi-\omega\right)} = -\cos{\phi}\sin{\omega} + \sin{\phi}\cos{\omega}$, together with the choice $\sqrt{2-a}\,\sin{\omega} = 1+\sqrt{a} - a$, and $\sqrt{2-a}\cos{\omega} = \sqrt{1-a}\left(1+\sqrt{a}\right)$ to simplify this to \begin{equation}
\langle CHSH \rangle_{max} \ge \left(1+\sqrt{a}-a\right) +\sqrt{2-a} \sin{\left(\phi - \omega\right)}
\end{equation}
For any $a \in [0,1)$ this is always greater than 2, provided Alice chooses $\bar{A}_1, \bar{A}_2 $ so that $\phi - \omega = \pi/2$.

Note that when  $a = 1$  (i.e. in the limiting case $\bra{\chi} U \ket{\chi} = 1$) the control unitary obviously generates no entanglement-- and consequently we do not expect any Bell inequality violation when $a=1$. This is not a problem for our application which should always fall in the regime $a < 1$.

Hence Alice and Bob should be able to violate a Bell inequality using the output of the circuits in Fig. \ref{fig3} and  \ref{fig2}. In addition Bob may uses $X$ and $Z$-directions as his two measurement settings during the Bell test. By using these settings we can establish the Z-outcomes of the control qubit in the contextuality protocol are not realistic. Consequently during any run of the contextuality test, the choice of which two observables are being measured on the register is not predetermined  (because this choice depends on the Z-outcome of the control qubit - a non realistic quantity).

\section{Appendix B: Even $n- \textrm{cycles}$ technical details}\label{sec:evenncyclestechnical}

$2m$-cycle contextuality tests, are usually performed on $4$-level quantum systems in the state $\ket{\psi} = 1/\sqrt{2} \ket{0} + 1/\sqrt{2} \ket{3} $, where we have chosen to label the Hilbert space basis $\{\ket{0},\ket{1},\ket{2}, \ket{3}\}$. The optimal settings for observables $j = 0, \dots, n-1$ are

\begin{eqnarray}\label{eq:4levelchainedbell}
&&A_{j} = \cos{\frac{\pi j}{n}}Z_1 + \sin{\frac{\pi j}{n}} X_1 \qquad \textrm{for even j and j = 0,}\notag\\
&&A_{j} = \cos{\frac{\pi j}{n}}Z_2 + \sin{\frac{\pi j}{n}} X_2 \qquad  \textrm{for odd j},
 \end{eqnarray}
 where any pair $A_j$ and $A_{j + 1 \, {\rm mod} \, n}$ are compatible and
 \begin{eqnarray}\label{red}
 X_1 &=& \sum_k \ket{k+ 2  \, {\rm mod} \, 4}\bra{k}, \, \,\, Z_1 = \sum_k (-1)^{\floor{k/2}}\ket{k}\bra{k}, \\  X_2 &=& \sum_k \ket{k + (-1)^k \, {\rm mod}\, 4}\bra{k} \textrm{ and } Z_2 = \sum_k (-1)^k \ket{k} \bra{k} \notag
 \end{eqnarray}

For example in the $4$-cycle case we have:
\begin{align}
A_0 = Z_1 && A_1 =  \frac{1}{\sqrt{2}}\left(X_2 + Z_2\right) \notag\\
A_2  =  X_1 && A_3 =    \frac{1}{\sqrt{2}} \left(X_2-Z_2\right)
\end{align}

  The theory of Euler rotations will give a systematic method for computing $U$, provided we use the representations of $A_0,\dots, A_{n-1}$ given in Refs. \cite{Braunstein,Araujo,Grudka}, we provide another Appendix on how this representation is related to \eqref{eq:4levelchainedbell}-\eqref{red}. All other details should follow unaltered.

\section{Appendix C: Odd $n- \textrm{cycles}$ technical details}

Tests of $2m+1$-cycle inequalities based on Eq. \eqref{eq:ncycles}, are typically implemented in a 3-level quantum system. The observables are:
\begin{equation}
A_j = 1- 2\ket{v_j}\bra{v_j}
\end{equation}
 where the ray
 \begin{equation}\label{eq:rays}
v_j~=~( \sin\phi \, \cos\frac{\pi j (n-1)}{n} ,\sin \phi \, \sin\frac{\pi j (n-1)}{n} , \cos\phi )
\end{equation}
for $\cos^2\phi = \cos\frac{\pi}{n} /(1+ \cos\frac{\pi}{n})$ \cite{Araujo} and $j \in \{0,\dots, n-1\}$. The state that violates this inequality maximally, and is therefore usually adopted in experimental tests is $\ket{\psi} = \ket{2}$.

The protocol could be implemented by control-unitary based:
\begin{equation}
U_{j,j+1} = -\frac{1}{\sqrt{2}}\ket{v_j}\bra{-_j} +  \frac{1}{\sqrt{2}}\ket{v_{j+ 1}}\bra{+_j} + \ket{v_{j\times j+1}}\bra{v_{j\times j+1}},
\end{equation}
 where $\{\ket{v_{j\times j+1}}, \ket{v_j}, \ket{v_{j+1}}\}$ form a basis for the 3-d Hilbert space, $\bra{-_j} = \bra{v_{j+1}} - \bra{v_j} $ and $\bra{+_j} = \bra{v_j} + \bra{v_{j+1}}$. This is tantamount to a $\pi/2$ rotation in the plane spanned by $\ket{v_j}$ and $\ket{v_{j+1}}$.

\section{Appendix D:  Chained Bell vs Even n-cycle inequalities}

All $n$-cycle inequalities for even $n$ have a one to one mapping onto a chained Bell inequalities (where Alice and Bob each have one half of a Bell pair and $n/2$ observables). In this case our 4-level system basis vectors $\{\ket{0}, \ket{1},\ket{2}, \ket{3}\}$ from Sec. \ref{sec:evenncycles} should be identified with a basis for the 2-qubit space according to $\ket{0} \leftrightarrow \ket{00}, \, \ket{1}\leftrightarrow \ket{01}, \ket{2}\leftrightarrow \ket{10}, \, \ket{3}\leftrightarrow \ket{11}$. The state that violates the chained Bell inequality maximally is $\frac{1}{\sqrt{2}} \ket{00} + \frac{1}{\sqrt{2}} \ket{11}$ and the optimal settings are
\begin{eqnarray}\label{eq:chianedbellsettings}
&&A_{i} = \left(\cos{\frac{2\pi i}{n}}Z + \sin{\frac{2\pi i}{n}} X\right) \otimes I, \notag\\
&&\textrm{for Alice's settings $i= 0,\dots, n/2-1$.}\notag\\
&&B_{i} = I \otimes \left(\cos{\frac{\pi (2i+1)}{n}}Z + \sin{\frac{\pi (2i+1)}{n}} X\right), \notag\\
&& \textrm{for Bob's settings $i= 0,\dots, n/2 -1$.}
 \end{eqnarray}
For an exact correspondence with Sec.  \ref{sec:evenncycles} note that with the basis identification above $X\otimes I =   X_1$ in Eq. \eqref{red}. Similarly $Z \otimes I = Z_1$ , while $I\otimes Z = Z_2$ and $I\otimes X = X_2$.  We highlight that for the CHSH inequality
\begin{equation}
\langle A_0 B_0 \rangle + \langle A_1B_0\rangle +\langle A_1 B_1\rangle - \langle A_0 B_1\rangle \le 2
\end{equation}
using the Bell state $\phi^+ = \frac{1}{\sqrt{2}}\ket{00} +\frac{1}{\sqrt{2}} \ket{11}$, the settings prescribed by \eqref{eq:chianedbellsettings} are
\begin{align}
A_0 = Z \otimes I \qquad &&  B_0 = I\otimes \frac{1}{\sqrt{2}} \left( Z + X \right)\notag\\
  A_1 = X \otimes I  \qquad && B_1 = I \otimes \frac{1}{\sqrt{2}}\left(X-Z\right)
 \end{align}
 The problem of finding $U$ such that $A_j = U^{\dagger} A_0 U$ and $B_j = U^{\dagger}B_{n/2-1}U$ in  Sec. \ref{sec:evenncycles}, now inheres a systematic solution from the theory of Euler rotations. This solution is the form $U = U_1\otimes U_2$, where each single qubit unitary  $U_i = e^{-i\theta_i \sigma \cdot r_i}$ for $i= 1,2$, has the free parameters, $\theta_i, r_i$ (rotation angle and axis). Parameters $\theta_1 \,\&\, r_1$ are fixed by $ (U_1^{\dagger} \otimes I ) \, A_0 \, ( U_1 \otimes I ) = A_j$, while $\theta_2 \,\& \,r_2$ are fixed by $ (I \otimes U_2^{\dagger})  \, B_{n/2-1} \, (I\otimes U_2 )= B_j$. From the form of $A_j$ and $B_j$ in Eq. \eqref{eq:chianedbellsettings} we observe that it is always possible to set $r_1 = r_2 = \hat{y}$. Once we have found $U$, it is possible, albeit convoluted to rewrite $U = U_1\otimes U_2$ in the basis $\{\ket{0}, \ket{1} \ket{2},\ket{3}\}$.
\end{document}